\documentclass[nofootinbib,prd,preprintnumbers,superscriptaddress,aps]{revtex4}
\usepackage[utf8]{inputenc}
\pdfoutput=1

\usepackage{caption} 
\usepackage{comment} 
\usepackage{graphicx}
\usepackage{epsfig}
\usepackage{bm}
\usepackage{amssymb}
\usepackage{float}
\usepackage{amsmath}
\usepackage{dcolumn}
\usepackage{cancel}
\usepackage[colorlinks]{hyperref}
\usepackage[usenames, dvipsnames]{color}

\hypersetup{
     breaklinks=true, 
    pdfstartview={FitH},  
    colorlinks=true, 
    linkcolor=blue,  
    citecolor=red,  
    filecolor=magenta,  
    urlcolor=blue, 
    anchorcolor=green,  
    linktocpage=true
}

\providecommand{\U}[1]{\protect\rule{.1in}{.1in}}

\newcommand{\be}{\begin{equation}}
\newcommand{\ee}{\end{equation}}

\newcommand{\mincir}{\raise
-3.truept\hbox{\rlap{\hbox{$\sim$}}\raise4.truept\hbox{$<$}\ }}
\newcommand{\magcir}{\raise
-3.truept\hbox{\rlap{\hbox{$\sim$}}\raise4.truept\hbox{$>$}\ }}

\ifx\pdfoutput\relax\let\pdfoutput=\undefined\fi
\newcount\msipdfoutput
\ifx\pdfoutput\undefined\else
\ifcase\pdfoutput\else
\ifx\paperwidth\undefined\else
\ifdim\paperheight=0pt\relax\else\pdfpageheight\paperheight\fi
\ifdim\paperwidth=0pt\relax\else\pdfpagewidth\paperwidth\fi
\fi\fi\fi

\hypersetup{colorlinks=true,
	breaklinks=true,
	pdfstartview=Fit,
	linkcolor=blue,
	citecolor=blue,
	urlcolor=blue}

\begin{document}
\title{Modified cosmology through generalized mass-to-horizon entropy: observational constraints from DESI DR2 BAO data}

\author{Giuseppe Gaetano Luciano}
\email{giuseppegaetano.luciano@udl.cat}
\affiliation{Departamento de Qu\'{\i}mica, F\'{\i}sica y Ciencias Ambientales y del Suelo, Escuela Polit\'ecnica Superior -- Lleida, Universidad de Lleida, Av. Jaume II, 69, 25001 Lleida, Spain}

\author{Andronikos Paliathanasis}
\email{anpaliat@phys.uoa.gr}
\affiliation{School for Data Science and Computational Thinking, Stellenbosch 
University,44 Banghoek Rd, Stellenbosch 7600, South Africa}
\affiliation{Centre for Space Research, North-West University, Potchefstroom 
2520, South Africa}
\affiliation{Departamento de Matem\`{a}ticas, Universidad Cat\`{o}lica del 
Norte, Avda. Angamos 0610, Casilla 1280 Antofagasta, Chile}
\affiliation{National Institute for Theoretical and Computational Sciences (NITheCS), South Africa}

\begin{abstract}
A generalized mass–to–horizon entropy has recently been proposed as an extension of the Bekenstein–Hawking area law, derived from a modified mass–horizon relation constructed to ensure consistency with the Clausius equation. Within the gravity–thermodynamics conjecture, this entropy formulation yields modified Friedmann equations, which recover the standard $\Lambda$CDM cosmology in the appropriate limit of the model’s two free parameters. In the present study, we constrain this framework using observations from Type~Ia supernovae (SNIa), cosmic chronometers (CC) and baryon acoustic oscillations (BAO, including DESI~DR2), together with the SH0ES distance–ladder prior on $H_0$, across four combinations of data sets. Although the extended entropic scenario yields a slightly better, or statistically comparable, fit to the data, model selection via the Akaike Information Criterion (AIC) mildly favors the cosmological constant as the dark energy candidate. Moreover, the $\Lambda$CDM limit lies within $\sim\!1\sigma$ of our constraints, indicating no significant deviation from standard cosmology with current data. 
\end{abstract}

\maketitle

\section{Introduction}
\label{Intro}

Astrophysical measurements from diverse sources - including luminosity distances of Type~Ia supernovae~\cite{SupernovaSearchTeam:1998fmf,SupernovaCosmologyProject:1998vns}, 
temperature fluctuations in the cosmic microwave background~\cite{COBE:1992syq,WMAP:2003ivt}, 
and large-scale structure surveys~\cite{2DFGRS:2001zay,SDSS:2003eyi,BOSS:2016wmc} -  
consistently indicate that the Universe has experienced two distinct accelerated expansion phases. 
The first corresponds to a primordial inflationary stage in the very early Universe, while the second is associated with the present epoch of late-time acceleration. 
Explaining these phenomena has motivated two broad classes of theoretical frameworks.  

One approach seeks to modify the underlying geometry of spacetime itself. 
By relaxing the strict reliance on Einstein’s original field equations and extending the Einstein–Hilbert action, 
this strategy gives rise to a wide spectrum of theories collectively referred to as modified gravity~\cite{Capozziello:2011et}. On the other hand, a conceptually different route retains general relativity as the gravitational theory but alters the matter sector, 
introducing additional dynamical components - such as scalar fields (e.g., the inflaton) or exotic dark energy fluids - to drive cosmic acceleration~\cite{Olive:1989nu,Bartolo:2004if,Copeland:2006wr,Cai:2009zp,DiValentino:2025sru}.  

Recently, increasing attention has been drawn to a third line of thought, which highlights a deep interplay between gravitational dynamics and thermodynamics~\cite{Jacobson:1995ab,Padmanabhan:2003gd,Padmanabhan:2009vy}.  
In this picture, the Universe is treated as an equilibrium thermodynamic system enclosed by the apparent horizon, and the gravitational field equations emerge naturally by applying the first law of thermodynamics to this boundary~\cite{Frolov:2002va,Cai:2005ra,Akbar:2006kj,Cai:2006rs}.  
Such a thermodynamic formulation is not restricted to general relativity but can be extended to various modified gravity theories, provided that the horizon entropy--area relation is appropriately generalized~\cite{Paranjape:2006ca,Akbar:2006er,Jamil:2009eb,Cai:2009ph}.

Within this setting, the concept of horizon entropy acquires additional relevance in the framework of entropic cosmology~\cite{Easson:2010av}, 
where thermodynamic principles are directly applied to model large-scale cosmic evolution.  
Here, the notion of entropic forces, originating from boundary terms in the Einstein–Hilbert action, 
is employed to account for the present accelerated expansion \cite{Easson:2010av}, thereby eliminating the need for a dark energy component in the standard cosmological model.

Over recent decades, a number of generalized entropy formalisms have been proposed as extensions of the semiclassical Bekenstein–Hawking expression.  
These arise either from non-extensive statistical mechanics or from quantum and gravitational considerations applied to the holographic horizon.  
Prominent examples include Rényi~\cite{renyi1961entropy}, Tsallis~\cite{Tsallis:1987eu,Tsallis:2009}, Sharma–Mittal~\cite{Sharma1975}, 
Kaniadakis~\cite{kaniadakis2001non,Kaniadakis:2002zz,Luciano:2024bco}, and Barrow entropy~\cite{Barrow:2020tzx} (see also \cite{hanel2011comprehensive} for an axiomatic classification).  
These generalized entropies reduce to the standard Bekenstein–Hawking form for specific parameter choices, 
and their incorporation into gravitational thermodynamics has been the subject of extensive research~\cite{Lymperis:2018iuz,Saridakis:2020lrg,Nojiri:2019skr,Hernandez-Almada:2021rjs,Dheepika:2022sio,Jizba:2022icu,Lambiase:2023ryq,Jizba:2024klq,Ebrahimi:2024zrk,Nojiri:2025gkq}.

A conceptual subtlety, however, has been emphasized in recent studies~\cite{Nojiri:2021czz,Gohar:2023lta}:  
if entropy is generalized, thermodynamic consistency - as dictated by the first law - may require simultaneous modifications of other thermodynamic quantities, 
such as the temperature or internal energy~\cite{Nojiri:2022sfd,Nojiri:2021czz}.  From a cosmological standpoint, it has been argued~\cite{Gohar:2023hnb,Gohar:2023lta} that when the Clausius relation is imposed to fix the horizon temperature, 
and a linear mass-to-horizon relation (MHR) is assumed,  
entropic force models become effectively indistinguishable from the standard Bekenstein–Hawking scenario.  
This degeneracy arises independently of the specific entropy functional adopted, 
implying that all such models inherit the same shortcomings as the conventional entropy–area law in explaining cosmic acceleration at both the background and perturbation levels~\cite{Basilakos:2012ra,Basilakos:2014tha}.  
To overcome this limitation, a generalized MHR has been proposed, yielding a two-parameter modified entropy expression \cite{Gohar:2023hnb,Gohar:2023lta}.

The cosmological viability of the generalized mass-to-horizon entropy framework has been examined in~\cite{Gohar:2023lta},  
where it was found that, for appropriate parameter values, the model matches observational constraints with accuracy comparable to that of the $\Lambda$CDM paradigm.  
Moreover, by implementing the gravity–thermodynamics correspondence,  
modified Friedmann equations have been derived in~\cite{Basilakos:2025wwu},  
revealing the emergence of an effective dark energy sector driven by the additional terms introduced through the generalized entropy.  

In the realm of observational cosmology, the recent data releases from the Dark Energy Spectroscopic Instrument (DESI), including its upgraded DESI~DR2, have already established themselves as powerful tools for placing stringent constraints on a wide range of cosmological models \cite{DESI:2025fii,DESI:2025zgx}. 
Examples include dynamical dark energy models~\cite{Ormondroyd:2025iaf,You:2025uon,Gu:2025xie,Santos:2025wiv,Li:2025cxn,Alfano:2025gie,Carloni:2024zpl,Luciano:2025elo,Luciano:2025hjn},  
early dark energy scenarios~\cite{Chaussidon:2025npr}, scalar field theories with either minimal or non-minimal couplings~\cite{Anchordoqui:2025fgz,Ye:2025ulq,Wolf:2025jed,Paliathanasis:2025xxm,Arora:2025msq},  
Generalized Uncertainty Principle models~\cite{Paliathanasis:2025dcr}, interacting dark sector frameworks~\cite{Shah:2025ayl,Silva:2025hxw,vanderWesthuizen:2025iam,Gumjudpai:2025tkc,Guedezounme:2025wav},  
astrophysical applications~\cite{Alfano:2024jqn}, cosmographic investigations~\cite{Luongo:2024fww},  
and various modified gravity proposals~\cite{Yang:2025mws,Paliathanasis:2025hjw,Tyagi:2025zov,Li:2025dwz,Sharma:2025qmv}.

Building upon the above premises, in this work we employ observational data to constrain the modified cosmological framework grounded in the generalized mass-to-horizon entropy.
The paper is structured as follows: in the next Section, we review the derivation of the modified Friedmann equations arising from the application of the gravity–thermodynamics conjecture within the framework of the generalized mass-to-horizon entropy.  
In Sec.~\ref{tests}, we present the parameter constraints for the entropic model derived from a joint analysis of DESI~DR2 BAO data, the Pantheon+ Type~Ia Supernova sample, and Cosmic Chronometer measurements.  
Section~\ref{Conc} contains a summary of the results along with our concluding remarks.  Unless otherwise stated, we adopt natural units throughout the theoretical analysis.

\section{Modified cosmology through generalized mass-to-horizon
entropy}
\label{ModCos}
We start our investigation by reviewing the gravity-thermodynamic conjecture in the  context of general relativity. This framework is subsequently extended by incorporating the mass-to-horizon relation (MHR) and the ensuing modified entropy introduced in \cite{Gohar:2023hnb,Gohar:2023lta}. Toward this end, we basically follow the treatment of \cite{Basilakos:2025wwu}.

We carry out our discussion within the setting of a spatially flat Friedmann–Lema\^{\i}tre–Robertson–Walker (FLRW) background, described by the metric
\be
ds^2 = g_{\mu\nu}dx^{\mu}dx^{\nu} = \ell_{\alpha\beta}  dx^\alpha dx^\beta  +  \tilde r^2\left(d\theta^2  +  \sin^2\theta\, d\phi^2\right)\, ,
\label{FRW}
\ee
where $\tilde r=a(t)\hspace{0.2mm}r$, $x^0=t$, $x^1=r$, $\ell_{\alpha\beta}=\mathrm{diag}\left(-1,a^2\right)$ and $a(t)$ is the time-dependent scale factor. Furthermore, we assume that the Universe is filled with a perfect fluid composed of both dust matter - encompassing cold dark matter and baryonic components - and radiation.

Within this theoretical framework, the dynamical apparent horizon assumes a fundamental role in characterizing thermodynamic quantities. In a spatially flat FLRW Universe, the radius of this horizon is expressed as $\tilde{r}_A = 1/H$~\cite{Frolov:2002va,Cai:2005ra,Cai:2009qf}, where the Hubble parameter $H = \dot{a}/a$ quantifies the expansion rate of the Universe, with the dot denoting differentiation with respect to cosmic time. The temperature associated with the apparent horizon is generally identified with a Hawking-like temperature~\cite{Hawking:1975vcx}, given by
\begin{equation}
\label{HaT}
   T_h = \frac{1}{2\pi \tilde{r}_A}\,,
\end{equation}
drawing a parallel with the thermodynamics of black holes~\cite{Cai:2009qf,Padmanabhan:2009vy}. 

In the present analysis, we adopt the assumption of a quasi-static cosmological evolution~\cite{Luciano:2023zrx}, which guarantees that the horizon temperature remains well-defined throughout cosmic history. Furthermore, we consider the cosmic fluid to be in thermal equilibrium with the apparent horizon, a condition supported by long-term interaction mechanisms acting over cosmological timescales~\cite{Padmanabhan:2009vy,Frolov:2002va,Cai:2005ra,Izquierdo:2005ku,Akbar:2006kj}. This equilibrium assumption enables the application of conventional thermodynamic laws and obviates the need for more complex non-equilibrium treatments.

The subsequent stage of the analysis involves attributing an entropy to the apparent horizon. In the context of general relativity, this is conventionally accomplished by employing the Bekenstein–Hawking entropy formula, originally proposed in the setting of black hole thermodynamics. According to this prescription, the entropy is given by $S_{BH} = A/4$, where $A = 4\pi \tilde{r}_A^2$ represents the surface area of the apparent horizon~\cite{Bekenstein:1973ur,Bekenstein:1974ax}.

Assuming that the Universe is permeated by a perfect fluid, the corresponding energy-momentum tensor adopts the canonical form
\begin{equation}
\label{cont}
T_{\mu\nu} = (\rho + p)\, u_{\mu} u_{\nu} + p\, g_{\mu\nu} \,,
\end{equation}
where \( \rho \) represents the energy density, \( p \) is the isotropic pressure and \( u^\mu \) denotes the four-velocity of the fluid elements. Within this framework, the conservation of energy and momentum, \( \nabla_\mu T^{\mu\nu} = 0 \), leads to the continuity equation
\begin{equation}
\dot{\rho} + 3H(\rho + p) = 0\,,
\end{equation}
which governs the temporal evolution of the energy density \( \rho \) in an expanding Universe. The corresponding work density, which emerges from changes in the apparent horizon radius, is defined as \( \mathcal{W} = -\frac{1}{2} \, \text{Tr}(T^{\mu\nu}) = \frac{1}{2}(\rho - p) \), where the trace is taken with respect to the induced metric on the \((t, r)\) submanifold, i.e., \( \text{Tr}(T^{\mu\nu}) = T^{\alpha\beta} h_{\alpha\beta} \).

Let us now recall that the gravity-thermodynamics conjecture posits that Einstein’s field equations may be interpreted as emerging from local thermodynamic identities applied to causal horizons. Within a cosmological setting, this perspective leads to the profound insight that the Friedmann equations can be derived from the application of the first law of thermodynamics to the apparent horizon. 

To illustrate this connection, let us consider the thermodynamic relation
\begin{equation}
dU = T_h\, dS - \mathcal{W}\, dV\,,
\label{14c}
\end{equation}
where \( dU \) denotes the infinitesimal variation of the internal energy of the Universe during an interval \( dt \), as a consequence of the change in the volume \( dV = 4\pi \tilde{r}_A^2\, d\tilde{r}_A \) bounded by the apparent horizon. Noting that this internal energy change corresponds to a decrease in the total energy \( E = \rho V \) contained within that volume, i.e., \( dU = -dE \), Eq.~\eqref{14c} can be reformulated to yield the second Friedmann equation,
\begin{equation}
\dot{H} = -4\pi \left(\rho + p\right)\,,
\label{F1}
\end{equation}
where the derivation assumes an adiabatic evolution of the apparent horizon~\cite{Cai:2005ra,Sheykhi:2018dpn,Luciano:2023zrx,Luciano:2025hjn}.

By substituting the continuity equation \eqref{cont} into Eq.~\eqref{F1} and integrating, one arrives at the first Friedmann equation
\begin{equation}
    H^2 = \frac{8\pi \rho}{3} + \frac{\Lambda}{3}\,,
    \label{F2}
\end{equation}
where the constant of integration $\Lambda$ is naturally interpreted as the cosmological constant.

Hence, the application of the gravity–thermodynamics conjecture at the cosmological horizon reproduces the standard Friedmann equations. As discussed above, extending this framework to include generalized entropy formalisms leads to modified versions of these equations. These modifications naturally give rise to alternative cosmological scenarios, wherein the additional entropic terms effectively mimic the behavior of a dark energy component~\cite{Lymperis:2018iuz,Saridakis:2020lrg,Nojiri:2019skr,Hernandez-Almada:2021rjs,Dheepika:2022sio,Jizba:2022icu,Lambiase:2023ryq,Jizba:2024klq,Ebrahimi:2024zrk,Nojiri:2025gkq}. In the subsequent analysis, we adopt the gravity–thermodynamic approach by incorporating the generalized mass-to-horizon entropy relation recently proposed in~\cite{Gohar:2023hnb,Gohar:2023lta}.

\subsection{Generalized mass-to-horizon relation and modified Cosmology}
\label{GIa}
As outlined in Sec.~\ref{Intro}, a generalized MHR was recently introduced in~\cite{Gohar:2023hnb,Gohar:2023lta} within the context of entropic cosmology and holographic-based scenarios. This proposal is primarily motivated by the recognition that, when the Clausius relation is imposed to maintain thermodynamic consistency and a linear MHR is assumed, the resulting entropic force on cosmological horizons remains identical to that obtained using the standard Bekenstein entropy and Hawking temperature, irrespective of the specific entropy formalism employed. Consequently, such models inevitably inherit the same limitations encountered in conventional Bekenstein–Hawking-based entropic cosmologies when attempting to account for the observed large-scale dynamics of the Universe~\cite{Basilakos:2012ra,Basilakos:2014tha}.

These considerations have led to the formulation of a generalized MHR in the form $M = \gamma \frac{c^2}{G} L^n$, 
where \( M \) denotes the effective mass associated with the system, \( L \) corresponds to the cosmological horizon, \( \gamma \) is a constant with dimensions $[L]^{1-n}$ and \( n \) is a non-negative real parameter (here, the fundamental constants \( c \) and \( G \) have been reinstated for consistency with the conventions used in~\cite{Gohar:2023hnb,Gohar:2023lta}).  
Notably, for specific values of the free parameters \( \gamma \) and \( n \), the resulting cosmological dynamics exhibit excellent agreement with current observational data sets. This consistency underscores the potential of the generalized MHR framework as a compelling alternative to the standard cosmological model, offering fresh theoretical insight into the origin and interpretation of the cosmological constant \cite{Gohar:2023hnb,Gohar:2023lta,Basilakos:2025wwu}.

By combining the generalized MHR with the Clausius relation, and employing the Hawking temperature defined in Eq.~\eqref{HaT}, one obtains a generalized entropy expression of the form~\cite{Gohar:2023hnb,Gohar:2023lta}
\begin{equation}
\label{GMHE}
    S = \gamma\,\frac{2n}{n+1}\,\tilde{r}_A^{n-1}\,S_{BH}\,,
\end{equation}
where \( S_{BH} \) represents the standard Bekenstein–Hawking entropy, and the apparent horizon \( \tilde{r}_A \) is adopted as the relevant length scale \( L \). It is readily verified that the conventional case, based on a linear MHR and the entropy–area law, is recovered for \( n = 1 \) and \( \gamma = 1 \). Since any physically viable deviations from the standard area-law are expected to be small, our analysis henceforth focuses on perturbative departures from this baseline configuration, i.e., $|n-1|, |\gamma-1|\ll1$.  This assumption is consistent with current observational constraints reported in~\cite{Gohar:2023lta, Basilakos:2025wwu}.

The above considerations have served as foundations for constructing a modified cosmological framework, as developed in Ref.~\cite{Basilakos:2025wwu}. Specifically, by applying the gravity–thermodynamics correspondence and following the same methodological steps outlined earlier, one obtains the modified Friedmann equations
\begin{eqnarray}
\label{FM1}
    H^2 &=& \frac{8\pi}{3} \left(\rho + \rho_{DE}\right), \\[2mm]
\label{FM2}
    \dot{H} &=& -4\pi \left(\rho + p + \rho_{DE} + p_{DE}\right),
\end{eqnarray}
where the influence of the generalized entropy \eqref{GMHE} manifests as an effective dark energy component, characterized by an energy density and pressure given by
\begin{eqnarray}
\label{rhode}
    \rho_{DE} &=& \frac{3}{8\pi} \left[ \frac{\Lambda}{3} + H^2 - \frac{2\gamma n}{3 - n}\, H^{3 - n} \right], \\[2mm]
\label{pde}
    p_{DE} &=& -\frac{1}{8\pi} \left[ \Lambda + \left(2\dot{H} + 3H^2\right) - 2\gamma n H^{1 - n} \left(\dot{H} + \frac{3}{3 - n} H^2 \right) \right], 
\end{eqnarray}
respectively.

It is straightforward to verify that the standard cosmological model is recovered in the special case \( n = \gamma = 1 \). Hence, the emergence of dark energy in this framework is not the result of explicitly introducing an additional component into the Universe’s energy content. Instead, it arises naturally as a consequence of the generalized relations introduced in Eq. \eqref{GMHE}.

The modified Friedmann equations derived above can be further reformulated by introducing the dimensionless fractional energy densities \( \Omega_i \equiv \dfrac{8\pi\rho_i}{3H^2} \), where the index \( i = m, r \) refers to pressureless matter and radiation, respectively, and \( \Omega_{DE} \equiv \dfrac{8\pi\rho_{DE}}{3H^2} \) denotes the effective dark energy contribution. From Eq.~\eqref{FM1}, one immediately obtains 
\begin{equation}
\label{cond1}
\Omega_{m} + \Omega_r + \Omega_{DE} = 1\,.
\end{equation}
Denoting the present-day energy densities of matter and radiation by \( \rho_{m0} \) and \( \rho_{r0} \), and adopting their standard scaling relations with the scale factor \( a \), namely \( \rho_m(a) = \rho_{m0} / a^3 \) and \( \rho_r(a) = \rho_{r0} / a^4 \), 
we find
\begin{equation}
\label{Hquad2}
    H^2(a) = \frac{H_0^2}{1 - \Omega_{DE}(a)}\left(\frac{\Omega_{m0}}{a^3} + \frac{\Omega_{r0}}{a^4} \right),
\end{equation}
where \( H_0 \) is the present-day Hubble parameter (throughout this work, we adopt the convention that the subscript “0” designates the present-day value of the corresponding physical quantity).

For practical purposes, it is convenient to recast the above equations in terms of the redshift \( z \), defined via \( 1 + z = 1/a \) (we set $a_0=1$). By substituting Eq.~\eqref{rhode} into Eq.~\eqref{Hquad2}, we then obtain 
\begin{equation}
\label{Omde}
\Omega_{DE}(z) = 1 - \left\{ \frac{3 - n}{2\gamma n} \left[ H_0 \sqrt{\Omega_{m0} + \Omega_{r0}(1 + z)} \right]^{n - 1} (1 + z)^{\frac{3(n - 1)}{2}} \left[ 1 + \frac{\Lambda}{3 H_0^2 [\Omega_{m0} + \Omega_{r0}(1 + z)] (1 + z)^3} \right] \right\}^{\frac{2}{n - 3}}.
\end{equation}
Evaluating this expression at the present epoch \( z = 0 \), and invoking the normalization condition \eqref{cond1}, one can solve for the cosmological constant in terms of the model parameters as
\begin{equation}
\label{Lambda}
\Lambda = \frac{6\gamma n}{3 - n}\, H_0^{3 - n} - 3H_0^2(\Omega_{m0} + \Omega_{r0})\,.
\end{equation}

The late-time cosmological behavior associated with the model~\eqref{Omde} has been recently investigated in Ref.~\cite{Basilakos:2025wwu}, showing that it successfully reproduces the standard thermal history of the Universe, characterized by a sequential transition from radiation to matter domination, followed by the emergence of a dark energy-dominated era. Furthermore, depending on the values of the entropic parameters, the effective dark energy equation-of-state parameter may exhibit distinct dynamical behaviors \cite{Basilakos:2025wwu}. 

\section{Observational Tests}
\label{tests}
In this section, we present the observational tests for the cosmological theory described in Eq.~(\ref{Hquad2}). 
To assess the viability of the mass-to-horizon entropy as a dark energy candidate, 
we employ a combination of late-time cosmological observations. 
In particular, we make use of the most recent Pantheon+ compilation of Type~Ia Supernovae, the observational Hubble parameter measurements $H(z)$ obtained through the cosmic chronometer (CC) method, and the latest Baryon Acoustic Oscillation (BAO) data from the Dark Energy Spectroscopic Instrument (DESI). 
The combination of these complementary probes enables a robust and model-independent test of the proposed scenario, 
significantly reducing parameter degeneracies. 
Further details regarding the observational data sets and the statistical analysis methodology are provided below:
\begin{itemize}
\item \textbf{Type Ia Supernovae (SNIa):}  
The Pantheon+ catalogue contains 1701 light curves corresponding to 1550 spectroscopically confirmed supernova events. These data provide the observed distance modulus, $\mu^{\mathrm{obs}}$, at specific redshifts in the range $10^{-3} < z < 2.27$~\cite{pan}. 
The theoretical distance modulus is given by  
\begin{equation}
\mu^{\mathrm{th}} = 5 \log_{10} D_{L} + 25\,,
\end{equation} 
where $D_{L}$ is the luminosity distance.  
For a spatially flat FLRW geometry, $D_{L}$ is related to the Hubble parameter $H(z)$ via $D_{L}(z) = (1+z) \int \frac{dz'}{H(z')}$. 
In our analysis, we employ the Pantheon+ data set both with and without the SH0ES Cepheid calibration.

\item \textbf{Cosmic Chronometers (CC):} 
This data set provides direct measurements of the Hubble parameter $H(z)$, obtained through the differential age method applied to cosmic chronometers, which are passively evolving galaxies with nearly synchronous stellar populations and similar evolutionary histories~\cite{Jimenez:2001gg,Moresco:2012by}.  
These measurements are model independent, relying solely on spectroscopic dating.  
In this work, we use the 31 $H(z)$ measurements reported in~\cite{cc1}, covering the redshift range $0.09 \leq z \leq 1.965$.

\item \textbf{Baryon Acoustic Oscillations (BAO):}  
We use the BAO measurements from the DESI DR2 release~\cite{DESI:2025zgx,DESI:2025fii}.
The data set provides observational estimates of:  
\begin{align}
\frac{D_{M}}{r_{d}} &= \frac{D_{L}}{(1+z) \, r_{d}}, \\[2mm]
\frac{D_{V}}{r_{d}} &= \frac{\left[D_{L} \, \frac{z}{H(z)} \right]^{1/3}}{r_{d}}, \\[2mm]
\frac{D_{H}}{r_{d}} &= \frac{1}{r_{d} \, H(z)},
\end{align}
where $D_{M}$ is the transverse comoving distance, $D_{V}$ is the volume-averaged distance, $D_{H}$ is the Hubble distance and $r_{d}$ is the sound horizon at the drag epoch.
\end{itemize}

To carry out our analysis, we consider four combinations of the aforementioned probes, optionally augmented by the SH0ES distance–ladder prior on $H_0$:
\[
\begin{aligned}
\mathbf{D}_{1} &:~ \mathrm{SN} + \mathrm{BAO},\\[2pt]
\mathbf{D}_{2} &:~ \mathrm{SN} + \mathrm{CC} + \mathrm{BAO},\\[2pt]
\mathbf{D}_{3} &:~ \mathrm{SN} + \mathrm{BAO} + \mathrm{SH0ES},\\[2pt]
\mathbf{D}_{4} &:~ \mathrm{SN} + \mathrm{CC} + \mathrm{BAO} + \mathrm{SH0ES}.
\end{aligned}
\]
For convenience, we summarize these data sets in Table~\ref{tab1}.\footnote{Here we restore the correct units, with $H_0$ expressed in $\mathrm{km\,s^{-1}\,Mpc^{-1}}$ and $r_d$ in Mpc. The unit of $\gamma$ is then determined accordingly, as discussed at the beginning of Sec.~\ref{GIa}.
}

\begin{table}[tbp] 
\centering

\begin{minipage}[t]{0.49\linewidth}
\centering
\captionof{table}{Definitions of Data sets used in the analysis}
\label{tab1}
\begin{tabular}{ccccc}\hline\hline
\textbf{Data} & $\mathbf{SN}$ & $\mathbf{CC}$ & $\mathbf{BAO}$ & $\mathbf{SH0ES}$\\\hline
$\mathbf{D}_{1}$ & $\checkmark$ & $\times$ & $\checkmark$ & $\times$\\
$\mathbf{D}_{2}$ & $\checkmark$ & $\checkmark$ & $\checkmark$ & $\times$\\
$\mathbf{D}_{3}$ & $\checkmark$ & $\times$ & $\checkmark$ & $\checkmark$\\
$\mathbf{D}_{4}$ & $\checkmark$ & $\checkmark$ & $\checkmark$ & $\checkmark$\\\hline\hline
\end{tabular}
\end{minipage}\hfill
\begin{minipage}[t]{0.49\linewidth}
\centering
\captionof{table}{Priors}
\label{tab2}
\begin{tabular}{cc}\hline\hline
\textbf{Priors of Free Parameters} & \\\hline
$\mathbf{H}_{0}$ & $\left[60,80\right]$\\[-1.45mm]
$\mathbf{\Omega}_{m0}$ & $\left[0.01,1\right]$\\[-1.45mm]
$\mathbf{n}$ & $\left[0.01,2.99\right]$\\[-1.45mm]
$\mathbf{\gamma}$ & $\left[0.01,3\right]$\\[-1.45mm]
$\mathbf{r}_{d}$ & $\left[130,160\right]$\\\hline\hline
\end{tabular}
\end{minipage}

\end{table}

We perform Bayesian parameter inference with \texttt{COBAYA}\footnote{\url{https://cobaya.readthedocs.io/}} using a custom theory module. 
The present-day radiation density parameter is fixed to the Planck 2018 value, 
\(\Omega_{r0}=0.415\,H_{0}^{-2}\). 
The priors on the free parameters \(\{H_{0},\,\Omega_{m0},\,n,\,\gamma,\,r_d\) are listed in Table~\ref{tab2}.

Furthermore, we apply the same observational tests as for the $\Lambda$CDM model. Since the two cosmologies have different numbers of free parameters, we compare their performance using the Akaike Information Criterion (AIC)~\cite{AIC}. 
For a maximum-likelihood value $\mathcal{L}_{\max}$ (equivalently $\chi^2_{\min}=-2\ln\mathcal{L}_{\max}$ for Gaussian errors), the AIC is
\begin{equation}
AIC\simeq-2\ln\mathcal{L}_{\max}+2\kappa.
\end{equation}
where $\kappa$ is the number of independently varied parameters. 
In our fits we take $\kappa=3$ for $\Lambda$CDM and $\kappa=5$ for the mass-to-horizon entropic model of Eq.~(\ref{Hquad2}). The difference of the $AIC$ parameters reads%
\begin{equation}
\Delta AIC=AIC_{\mathrm{MHR}}-AIC_{\Lambda}=-2\ln\frac{\mathcal{L}_{\max}%
^{\mathrm{MHR}}}{\mathcal{L}_{\max}^{\Lambda}}+4\,,
\end{equation}
where the label $\mathrm{MHR}$ refers to the generalized mass-to-horizon relation model.

According to the Akaike scale, if $|\Delta\mathrm{AIC}|<2$ the two models are statistically indistinguishable. For $2\le|\Delta\mathrm{AIC}|<6$ there is weak evidence favoring the model with the lower AIC; for $6\le|\Delta\mathrm{AIC}|<10$ the evidence is strong; and for $|\Delta\mathrm{AIC}|\ge 10$ there is decisive evidence in favor of the model with the lower AIC.
\subsection{Results}

For the cosmological theory (\ref{Hquad2}) and the data set $\mathbf{D}_{1}$,  
we obtain the best-fit parameters: $H_{0} = 69.8^{+4.3}_{-4.3}$, $\Omega_{m0} > 0.531$, $n = 0.923^{+0.075}_{-0.066}$ and $\gamma = 1.59^{+0.65}_{-0.65}$.  
The comparison with $\Lambda$CDM yields $\mathbf{\chi}_{\mathrm{MHR}\,\min}^{2} - \mathbf{\chi}_{\Lambda\min}^{2} = -0.5$ and $\mathbf{AIC}_{\mathrm{MHR}} - \mathbf{AIC}_{\Lambda} = +3.5$.  

For data set $\mathbf{D}_{2}$, the best-fit parameters are $H_{0} = 69.2^{+1.6}_{-1.6}$, $\Omega_{m0} > 0.530$, $n = 0.928^{+0.068}_{-0.068}$, and $\gamma = 1.60^{+0.66}_{-0.66}$.  
The corresponding comparison with $\Lambda$CDM gives $\mathbf{\chi}_{\mathrm{MHR}\,\min}^{2} - \mathbf{\chi}_{\Lambda\min}^{2} = -0.6$ and $\mathbf{AIC}_{\mathrm{MHR}} - \mathbf{AIC}_{\Lambda} = +3.4$.  

Using data set $\mathbf{D}_{3}$, we find $H_{0} = 73.6^{+1.0}_{-1.0}$, $\Omega_{m0} > 0.530$, $n = 0.923^{+0.069}_{-0.069}$, and $\gamma = 1.57^{+0.67}_{-0.67}$.  
The comparison with $\Lambda$CDM yields $\mathbf{\chi}_{\mathrm{MHR}\,\min}^{2} - \mathbf{\chi}_{\Lambda\min}^{2} = -0.6$ and $\mathbf{AIC}_{\mathrm{MHR}} - \mathbf{AIC}_{\Lambda} = +3.4$. 

Finally, for data set $\mathbf{D}_{4}$, we obtain $H_{0} = 72.28^{+0.90}_{-0.90}$, $\Omega_{m0} > 0.519$, $n = 0.945^{+0.070}_{-0.070}$, and $\gamma = 1.70^{+0.86}_{-0.67}$.  
The statistical comparison with $\Lambda$CDM gives $\mathbf{\mathrm{MHR}\,\chi}_{\min}^{2} - \mathbf{\chi}_{\Lambda\min}^{2} = +0.0$ and $\mathbf{AIC}_{\mathrm{MHR}} - \mathbf{AIC}_{\Lambda} = +4.0$.  
These results are summarized in Table~\ref{tab3}. 

Based on our results, we conclude that the generalized entropic model provides a slightly better, or statistically comparable, fit to all four datasets when compared with the $\Lambda$CDM model. However, due to its additional degrees of freedom, the AIC mildly favors the cosmological constant as the preferred dark energy candidate. Notably, the limit $n = \gamma = 1$, corresponding to standard cosmology, lies within approximately $1\sigma$ of our constraints.

\begin{figure}[t]
\centering\includegraphics[width=0.75\textwidth]{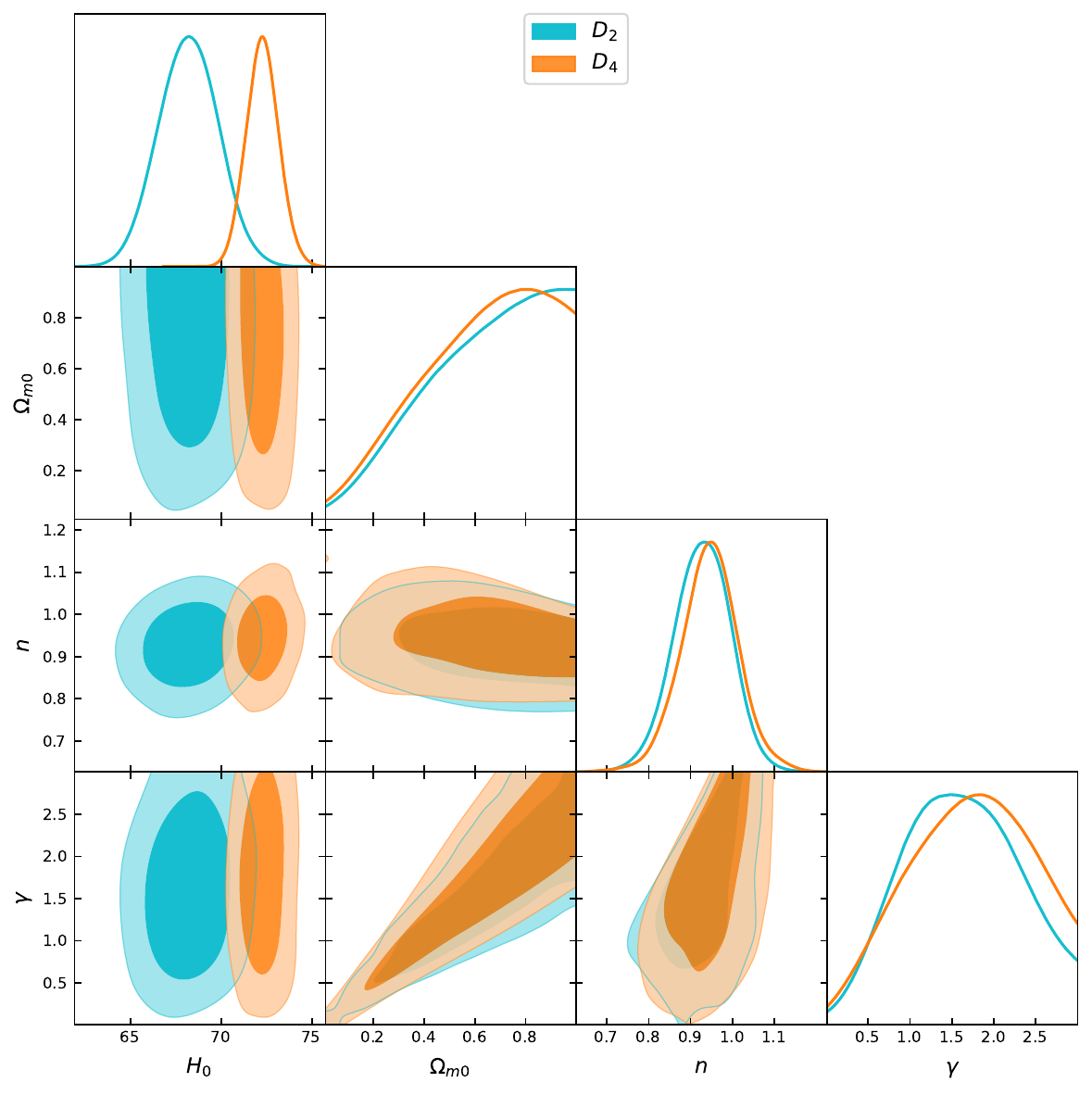}\caption{{
{Likelihood contours for the best-fit parameters of the generalized entropic model. We considered the data sets
$\mathbf{D}_{2}$  and $\mathbf{D}_{4}$, respectively.}}}%
\label{ff1}%
\end{figure}

In Fig.~\ref{ff1}, we show the $1\sigma$ and $2\sigma$ confidence contours for the best-fit parameters of model~(\ref{Hquad2}), as obtained from data sets $D_{2}$ and $D_{4}$. The plots were produced using the \texttt{GetDist} library\footnote{\url{https://getdist.readthedocs.io/}}~\cite{Lewis:2019xzd}.
It is particularly interesting that the $H_0$ estimate derived from the $\mathbf{D}_{4}$ dataset suggests a possible alleviation of the $H_0$ tension within the extended entropic framework~\cite{DiValentino:2025sru}. Nevertheless, we emphasize that our analysis does not incorporate CMB observations. This omission is critical when interpreting the apparent reduction in the Hubble tension, as our constraints rely solely on late-time and direct measurements, which typically yield higher $H_0$ values. In contrast, including early-universe and indirect probes, such as the CMB, generally leads to lower $H_0$ estimates (see, e.g.,~\cite{DiValentino:2021izs}).
Despite these caveats, our findings reveal intriguing features of the generalized entropic model that merit further investigation in future extensions of this work.

\begin{table}[tbp] \centering
\caption{Best-Fit Parameters}%
\begin{tabular}
[c]{ccccc}\hline\hline
\textbf{Best Fit}/\textbf{Data set} & $\mathbf{D}_{1}$ & $\mathbf{D}_{2}$ &
$\mathbf{D}_{3}$ & $\mathbf{D}_{4}$\\\hline
$\mathbf{H}_{0}$ & $69.8_{-4.3}^{+4.3}$ & $69.2_{-1.6}^{+1.6}$ &
$73.6_{-1.0}^{+1.0}$ & $72.28_{-0.90}^{+0.90}$\\
$\mathbf{\Omega}_{m0}$ & $>0.531$ & $>0.530$ & $>0.530$ & $>0.519$\\
$\mathbf{n}$ & $0.923_{-0.066}^{+0.075}$ & $0.928_{-0.068}^{+0.068}$ &
$0.923_{-0.069}^{+0.069}$ & $0.945_{-0.070}^{+0.070}$\\
$\mathbf{\gamma}$ & $1.59_{-0.65}^{+0.65}$ & $1.60_{-0.66}^{+0.66}$ &
$1.57_{-0.67}^{+0.67}$ & $1.70_{-0.67}^{+0.86}$\\
$\mathbf{\chi}_{\mathrm{MHR}\,\min}^{2}$ & $1419.4$ & $1434.8$ & $1469.5$ & $1492.1$\\
$\mathbf{\chi}_{\mathrm{MHR}\,\min}^{2}\mathbf{-\chi}_{\Lambda\min}^{2}$ & $-0.5$ & $-0.6$ &
$-0.6$ & $\,+0.0$\\
$\mathbf{AIC_{\mathrm{MHR}}-AIC}_{\Lambda}$ & $+3.5$ & $+3.4$ & $+3.4$ & $+4.0$\\\hline\hline
\vspace{2mm}
\end{tabular}
\label{tab3}%
\end{table}%

\section{Conclusions and Outlook}
\label{Conc}
In this work, we investigated a cosmological framework motivated by the proposed connection between thermodynamic principles and gravitational dynamics. It is well established that applying the first law of thermodynamics at the apparent horizon of the FRW Universe can reproduce the Friedmann equations. In the standard approach, this connection is formulated using the Bekenstein--Hawking entropy together with the Hawking temperature. We showed that extending the entropy expression beyond its conventional form leads, through the same thermodynamic reasoning, to modified Friedmann equations and, consequently, to new cosmological scenarios. 
In particular, we considered a recent generalization in which the relation between a horizon’s mass and its radius is modified, resulting in a generalized entropy characterized by two free parameters, $\gamma$ and $n$~\cite{Gohar:2023hnb,Gohar:2023lta}. 
The corresponding Friedmann equations, Eqs.~\eqref{FM1} and \eqref{FM2}, extend the $\Lambda$CDM paradigm, which is recovered as a special case in the limit $n = \gamma = 1$.

We used observational data from SNIa, CC and BAO, including the 
recently released DESI  DR2 data, in order to extract constraints on the 
scenario of modified cosmology.  As we showed, the best-fit value for the entropic exponent $n$ is found to be less than unity, whereas the corresponding estimate for $\gamma$ exceeds unity (for a comparison with other constraints, see \cite{Basilakos:2025wwu,Gohar:2023lta}). Notably, the standard cosmological model lies within approximately $1\sigma$ of our bounds for the datasets $\mathbf{D}_{1}:\mathrm{SN} + \mathrm{BAO}$, $\mathbf{D}_{2}: \mathrm{SN} + \mathrm{CC} + \mathrm{BAO}$, $\mathbf{D}_{3}: \mathrm{SN} + \mathrm{BAO} + \mathrm{SH0ES}$ and $\mathbf{D}_{4}: \mathrm{SN} + \mathrm{CC} + \mathrm{BAO} + \mathrm{SH0ES}$. Although the generalized entropic model yields a fit that is slightly better, or statistically comparable, to that of the $\Lambda$CDM model, its additional degrees of freedom lead the AIC to mildly favor the cosmological constant as the preferred dark energy candidate. Furthermore, for the case of $\mathbf{D}_{4}$, we obtained the value $H_0 = 72.28^{+0.90}_{-0.90}\,\mathrm{km\,s^{-1}\,Mpc^{-1}}$, which
may potentially alleviate the Hubble tension. Clearly, a conclusive evaluation will require a dedicated observational study, incorporating comparisons with data from various cosmological surveys, including Planck \cite{Planck:2018vyg} and H0LiCOW \cite{H0LiCOW:2019pvv}, which offer complementary determinations of the Hubble constant across distinct redshift ranges. An investigation in this direction is planned as a continuation of the present work.

In conclusion, the modified cosmological scenario emerging from the gravity–thermodynamics conjecture, when formulated with a generalized mass-to-horizon entropy, gives rise to non-trivial phenomenological implications. Nevertheless, further work is required before it can be regarded as a viable description of the Universe. In particular, a thorough perturbative analysis is needed to examine the evolution of matter overdensities and the growth of cosmic structures. Moreover, a comprehensive dynamical-systems investigation would be essential to uncover the global cosmic behavior in a model-independent manner. These important tasks fall beyond the scope of the present study and are deferred to future research.

\acknowledgments 
The research of GGL is supported by the postdoctoral fellowship program of the 
University of Lleida. GGL gratefully acknowledges  the contribution of 
the LISA 
Cosmology Working Group (CosWG), as well as support from the COST Actions 
CA21136 - \textit{Addressing observational tensions in cosmology with 
systematics and fundamental physics (CosmoVerse)} - CA23130, \textit{Bridging 
high and low energies in search of quantum gravity (BridgeQG)} and CA21106 - 
\textit{COSMIC WISPers in the Dark Universe: Theory, astrophysics and 
experiments (CosmicWISPers)}. AP was Financially supported by FONDECYT 1240514 ETAPA 2025.

\bibliography{Bib}

\end{document}